\documentclass[aps,prd,reprint,preprintnumbers,superscriptaddress,nofootinbib,longbibliography,floatfix]{revtex4-2}
\pdfoutput=1

\usepackage[utf8]{inputenc} 
\usepackage[T1]{fontenc}    
\usepackage{hyperref}       
\usepackage{url}            
\usepackage{booktabs}       
\usepackage{amsfonts}       
\usepackage{nicefrac}       
\usepackage{microtype}      
\usepackage{xcolor}         
\usepackage{amsmath}
\usepackage{graphicx}
\usepackage{float}
\usepackage{subfig}
\usepackage{amsmath}
\usepackage{physics}
\usepackage{xcolor}
\usepackage{multirow}

\hypersetup{
  colorlinks=true,
  citecolor=blue,
  linkcolor=blue,
  urlcolor=blue
}

\usepackage{physics-defs}
\usepackage{placeins}

\DeclareMathOperator*{\argmin}{arg\,min}

\newcommand{\A}{\mathcal{A}}

\newcommand{\model}{Covariant Particle Transformer}
\newcommand{\attn}{covariant attention}

\renewcommand{\eta}{y}
\usepackage{physics-defs}

\begin{document}

\title{A Holistic Approach to Predicting Top Quark Kinematic \\ Properties with the Covariant Particle Transformer}

\author{Shikai Qiu}
\email{calvin_qiu@berkeley.edu}
\affiliation{Department of Physics, University of California, Berkeley, Berkeley, CA 94720, USA}

\author{Shuo Han}
\email{shuohan@lbl.gov}
\affiliation{Physics Division, Lawrence Berkeley National Laboratory, Berkeley, CA 94720, USA}

\author{Xiangyang Ju}
\email{xju@lbl.gov}
\affiliation{Physics Division, Lawrence Berkeley National Laboratory, Berkeley, CA 94720, USA}

\author{Benjamin Nachman}
\email{bpnachman@lbl.gov}
\affiliation{Physics Division, Lawrence Berkeley National Laboratory, Berkeley, CA 94720, USA}
\affiliation{Berkeley Institute for Data Science, University of California, Berkeley, CA 94720, USA}

\author{Haichen Wang}
\email{haichenwang@berkeley.edu}
\affiliation{Physics Division, Lawrence Berkeley National Laboratory, Berkeley, CA 94720, USA}
\affiliation{Department of Physics, University of California, Berkeley, Berkeley, CA 94720, USA}

\begin{abstract}

Precise reconstruction of top quark properties is a challenging task at the Large Hadron Collider due to combinatorial backgrounds and missing information.  We introduce a physics-informed neural network architecture called the Covariant Particle Transformer (CPT) for directly predicting the top quark kinematic properties from reconstructed final state objects.  This approach is permutation invariant and partially Lorentz covariant and can account for a variable number of input objects.  In contrast to previous machine learning-based reconstruction methods, CPT is able to predict top quark four-momenta regardless of the jet multiplicity in the event.  Using simulations, we show that the CPT performs favorably compared with other machine learning top quark reconstruction approaches. We make our code available at \href{https://github.com/shikaiqiu/Covariant-Particle-Transformer}{\texttt{https://github.com/shikaiqiu/Covariant-Particle-Transformer}}.

\end{abstract}

\maketitle

\clearpage

\clearpage

\section{Introduction}

For the Large Hadron Collider (LHC) experiments, the kinematic reconstruction of top quarks is critical to many precision tests of the Standard Model (SM) as well as direct searches for physics beyond the SM. Once produced, the top quark decays to a bottom quark ($b$-quark) and a W boson, with a branching ratio close to 100\%~\cite{10.1093/ptep/ptaa104}. Subsequently, the W boson decays into a lepton or quark pair. In the final state, quarks originating from top quark decays and other colored partons hadronize, resulting in collimated sprays of hadrons, known as jets. Conventional top quark methods assume that a hadronically decaying top quark produces three jets in the final state. Therefore, these methods are tuned to identify triplets of jets, which are considered as proxies for the three quarks originating directly from the top quark and W boson decays.  The estimated top quark four-momentum is computed from the sum of measured four-momenta over the triplet of jets. Essentially, top quark reconstruction is treated as a combinatorial problem of sorting jets, and most methods use jet kinematic and flavor tagging information to construct likelihood-based~\cite{Erdmann:2013rxa} or machine learning-based~\cite{ATLAS:2017fak, CMS:2019eih,Erdmann:2019evj, ATLAS:2020ior,Fenton:2020woz, Lee:2020qil, Shmakov:2021qdz,Badea:2022dzb} metrics to identify triplets of jets as proxies to top quarks and similar particles.

While the conventional top quark reconstruction approaches have been implemented in a variety of forms and extensively used at hadron collider experiments, they have fundamental flaws and shortcomings. The one-to-one correspondence between a parton (quark or gluon) and a jet, assumed by the conventional approaches, is only an approximation. Partons carry color charges but jets only consist of colorless hadrons. The formation of a jet, by construction, has to be contributed to by multiple partons. On the other hand, a single parton may contribute to the formation of multiple jets, particularly when the parton is highly energetic. In addition, triplet-based top quark reconstruction requires the presence of a certain number of jets in the final state. This jet multiplicity requirement can be inefficient because of kinematic thresholds, limited detector coverage, or the merging of highly collimated parton showers.

In this paper, we propose a new machine learning-enabled approach to determine the top quark properties through a holistic processing of the event final state. Our goal is to predict top quark four-momenta in a collision event with a given number of top quarks.  The number of top quarks can itself be learned from the final state or it can be posited for a given hypothesis.  As discussed earlier, the kinematic information of a top quark is not localized in 
a triplet of jets, rather, it is possessed by all particles in the event collectively. This motivates the use of particle identification (ID) and kinematic information from all detectable particles in the final state as input to the determination of the top quark four-momenta.   Specifically, the four-momenta and ID of all detectable final state particles are input to a deep neural networks regression model, which is constructed and trained to predict the four-momenta of a given number of top quarks. This approach offers three major advantages compared to conventional approaches. First, we no longer deal with the conceptually ill-defined jet-triplet identification process. Second, we can account for noisy or missing observations due to limited acceptance, detector inefficiency and resolution, as the regression model can learn such effects from Monte Carlo (MC) simulations. Third, the holistic processing of the event final state offers a unified approach to determining the top quark properties for both the hadronic and semi-leptonic top quark decays, which may simplify analysis workflows. Finally, our approach has a runtime polynomial in the number of final state objects as opposed to super-exponential for standard reconstruction-based approaches which need to consider all possible permutations, making ours the first tractable method for processes with high multiplicity final state such as \tttt.

To realize the holistic approach of top quark property determination, we propose a physics-informed transformer~\cite{vaswani2017attention} architecture termed Covariant Particle Transformer (CPT). CPT takes as input properties of the final state objects in a collision event and outputs predictions for the top quark kinematic properties.  Like other recent top reconstruction proposals~\cite{Fenton:2020woz, Shmakov:2021qdz, Lee:2020qil}, CPT is permutation invariant under exchange of the inputs. A novel attention mechanism~\cite{vaswani2017attention, bahdanau2014neural}, referred to as covariant attention, is designed to learn the predicted kinematic properties as a function of the set of final state objects as a whole, and guarantees that the predictions transform covariantly under rotation and/or boosts of the event along the beamline. While not fully Lorentz-covariant like Ref.~\cite{Bogatskiy:2020tje}, our approach captures the most important covariances relevant to hadron collider physics with minimal computational overhead and enjoys a much simpler implementation, which allows it to be easily adopted for a broad range of tasks in collider physics.

This paper is organized as follows.  Section~\ref{sec:model} introduces the construction and properties of CPT.  Synthetic datasets used for demonstrating the performance of CPT are introduced in Sec.~\ref{sec:data}.  Numerical results illustrating the performance of CPT are presented in Sec.~\ref{sec:perf}.  In Sec.~\ref{sec:ablat}, we explore what aspects of CPT give raise to the excellent performance.  The paper ends with conclusions and outlook in Sec.~\ref{sec:concl}.

\section{\model}
\label{sec:model}
\subsection{Symmetries and covariance}
At the LHC, the beamline determines a special direction and reduces the relevant symmetry group of collision events from the proper orthochronous Lorentz group $\mathrm{SO}^+(1, 3)$ to $\mathrm{SO}(2) \times \mathrm{SO}^+(1, 1) ,$ which contains products of azimuthal rotations and longitudinal boosts along the beamline. The \model\ extends the original transformer architecture to properly account for these symmetry transformations, by ensuring that if the four-momenta of all final state objects undergo such a transformation, the resulting prediction of the top quark four-momenta will undergo the same transformation. At its core, this is achieved through the novel covariant attention mechanism, which modifies the standard attention mechanism to ensure that all intermediate learned features have well-defined transformation properties.

Covariance\footnote{Called \textit{equivariance} in machine learning.} under rotations and boosts~\cite{Bogatskiy:2020tje,Shimmin:2021pkm} and input permutations~\cite{Dolan:2020qkr} have been studied in a variety of recent High Energy Physics (HEP) papers. A number of additional studies have explored permutation invariant architectures~\cite{Komiske:2018cqr,Qu:2019gqs,Moreno:2019bmu,Mikuni:2020wpr,Mikuni:2021pou} (see also other graph network approaches~\cite{Shlomi:2020gdn,2102.02770}). Compared to prior works in this direction, we make the following important contributions:
\begin{itemize}
    \item We develop the first transformer architecture that enforces Lorentz covariance. Transformers are a powerful class of neural networks that have revolutionized many areas of machine learning applications, such as natural language processing~\cite{vaswani2017attention, brown2020language}, computer vision~\cite{dosovitskiy2020image}, and recently protein folding~\cite{jumper2021highly}. By integrating the transformer architecture with Lorentz covariance, CPT combines the current state-of-the-art of machine learning with physics-specific knowledge to become a powerful tool for applications in collider physics, as we will illustrate in this work.
    \item We develop a simple, efficient, and effective way of achieving partial Lorentz covariance. While previous works have developed Lorentz covariant neural networks using customized architectures, they incur significant computational overhead compared to a standard neural network due to computations of continuous group convolutions~\cite{Shimmin:2021pkm} or irreducible representations of the Lorentz group~\cite{Bogatskiy:2020tje}. By contrast, CPT only requires a simple modification to the standard attention mechanism with minimal computational overhead.
    \item We are the first to demonstrate the benefit of using a Lorentz covariant architecture for regression problems where the targets are four momenta of the particles. Previous works on Lorentz covariant neural networks only evaluate on classification problems such as jet-tagging where the Lorentz group acts trivially (i.e. as an identity) on the targets. There Lorentz symmetry plays a less significant role since the neural network only needs to be Lorentz invariant but not covariant.
\end{itemize}

\subsection{Architecture}
The~\model\ consists of an encoder and a decoder. To ensure permutation invariance, we remove the positional encoding~\cite{vaswani2017attention} in the original transformer encoder. The encoder produces learned features of the final state objects, which include jets, photons, electrons, muons, and missing transverse energy (\met)\footnote{\met\ is implemented as a massless particle with zero longitudinal momentum component.}. 

\begin{figure}
    \centering
    \subfloat[CPT]{
    \includegraphics[width=0.32\columnwidth]{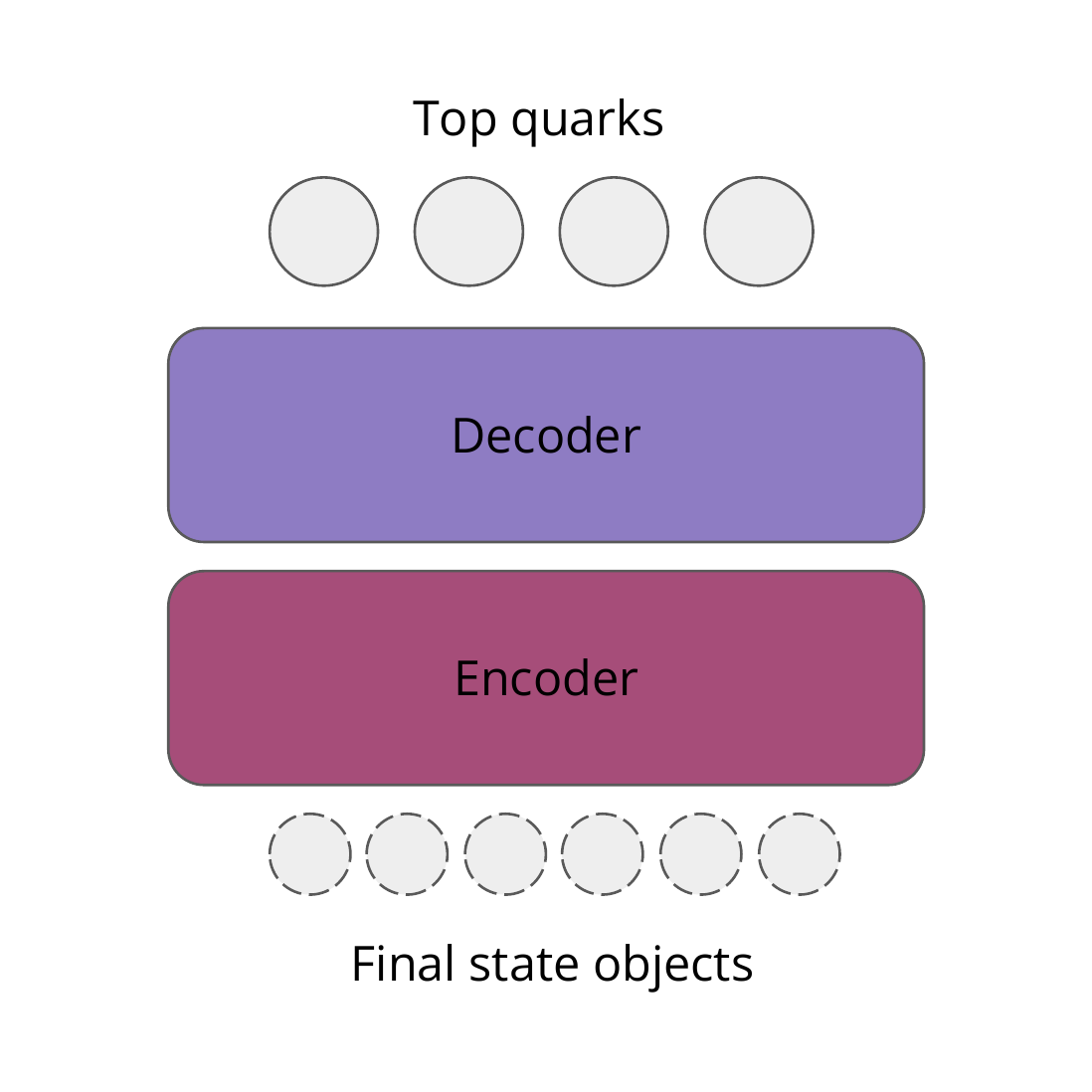}
    }
    \subfloat[Encoder]{\includegraphics[width=0.34\columnwidth]{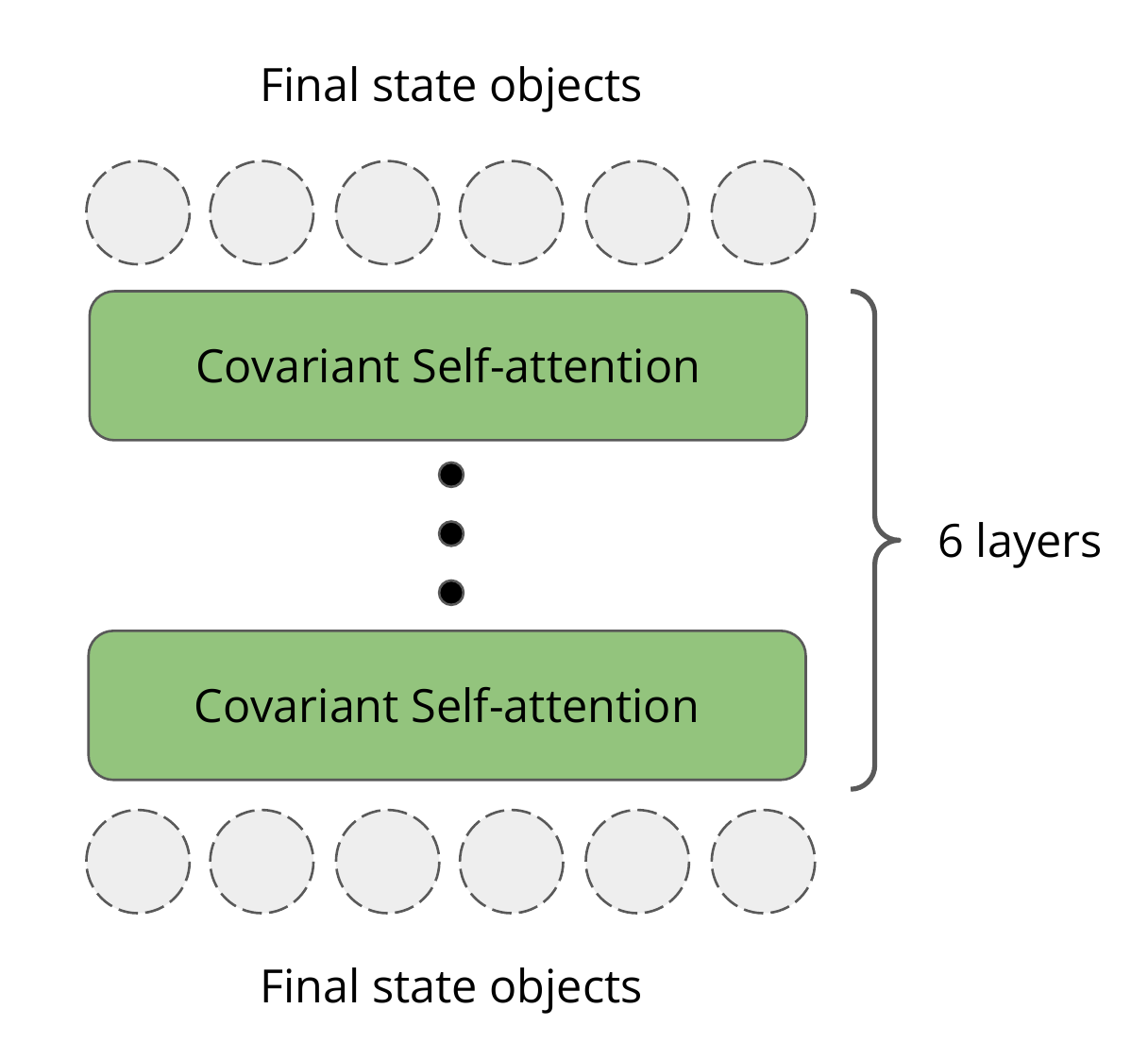}}
    \subfloat[Decoder]{\includegraphics[width=0.35\columnwidth]{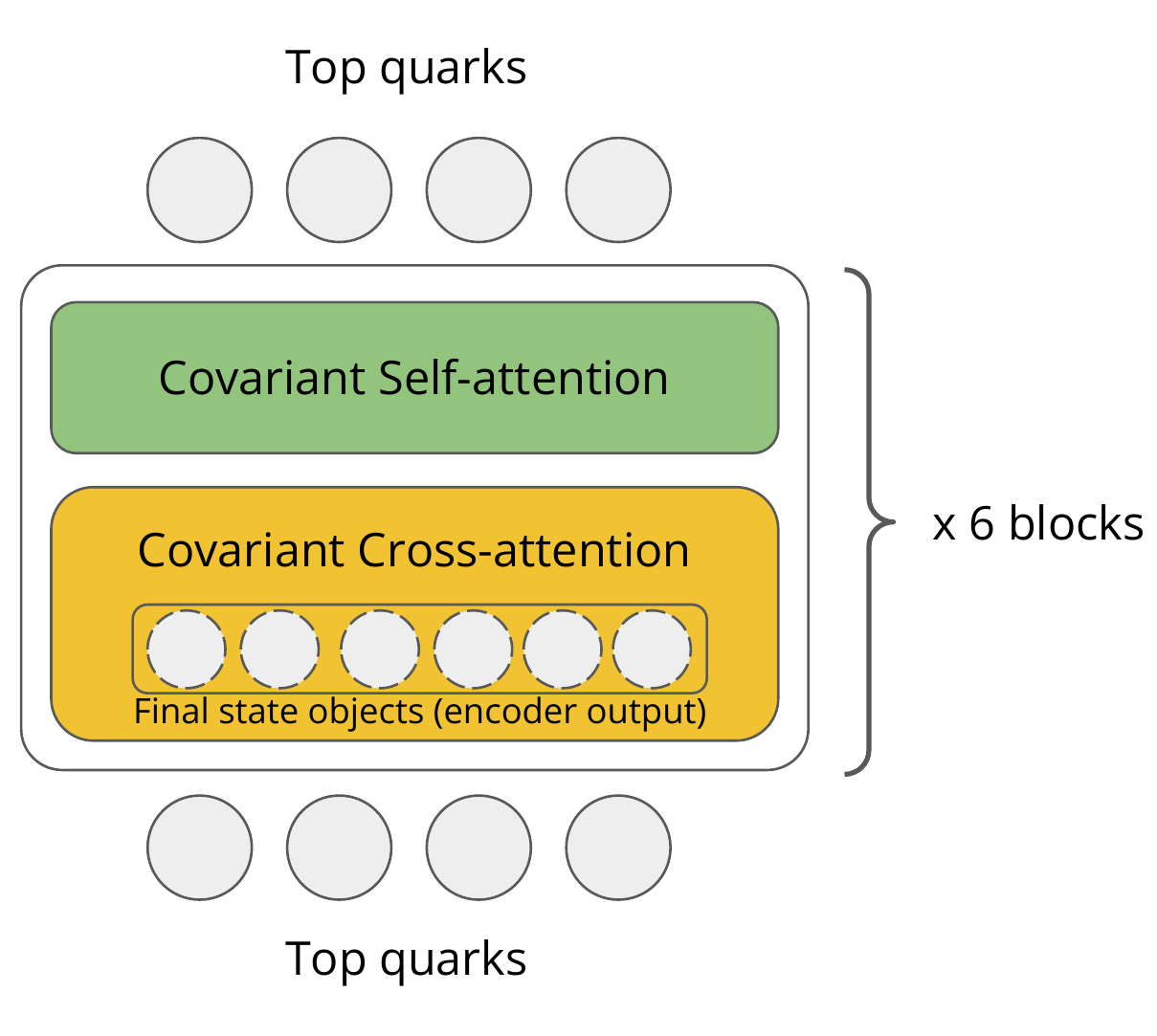}}
    \caption{An illustration of the Covariant Particle Transformer (CPT) architecture. The encoder consists of six covariant self-attention layers, while the decoder consists of six covariant cross-attention  layers and six covariant self-attention  layers interleaved. }
    \label{fig:architecture}
\end{figure}

\textcolor{black}{Each object is represented by its transverse momentum \pt, rapidity $\eta$, azimuthal angle $\phi$ expressed as a unit vector $(\cos(\phi), \sin(\phi))$ to avoid $\text{mod } \pi$ calculations, mass $m$, and particle identification ID.} The encoder uses six covariant self-attention layers to update the feature vectors of the final state objects. The decoder uses 12 \attn\ layers to produce learned features of the top quarks. Six of these layers use self-attention, which updates the feature vector of each top quark as a function of itself and the feature vectors of other top quarks, and the other six layers use cross-attention, which updates the feature vector of each top quark as a function of itself and the feature vectors of the final state objects. Finally, the feature vectors of top quarks are converted to predicted physics variables, which are the top quark four-momenta expressed in transverse momentum \pt, rapidity $\eta$, azimuthal angle unit vector, and mass $m$. Figure~\ref{fig:architecture} illustrates the architecture of the \model. Detailed descriptions of input featurization, CPT architecture, and the covariant attention mechanism are provided in Appendix~\ref{app:cpt_implementation}.

\subsection{Loss function}
The model is trained to minimize a supervised learning objective that measures the distance between the true and predicted values of the target variables\footnote{Note that learning the true value from reconstructed quantities introduces a prior dependence~\cite{ATL-PHYS-PUB-2018-013}.  This is true for nearly all regression approaches in HEP.}. Auxiliary losses are included to stabilize training the model. We provide a detailed description of the loss function in Appendix~\ref{sec:loss}.

\section{Datasets}
\label{sec:data}
We use Madgraph@NLO (v2.3.7)~\cite{Alwall:2014hca} to generate $pp$ collision events at next-to-leading order (NLO) in QCD. The decays of top quarks and W bosons are performed by MadSpin~\cite{Artoisenet:2012st}. We generate 9.2 million $t\bar{t}H$ events, 5.4 million $t\bar{t}t\bar{t}$ events, 1.3 million $t\bar{t}$ events, 1.3 million $t\bar{t}W$ events, and 1 million $t\bar{t}H$ events with a CP-odd top-Yukawa coupling (\ttHodd). In our generation, Higgs bosons decay through the diphoton channel for simplicity and all other objects such as top quarks and $W$ bosons decay inclusively.The Higgs Characterization model~\cite{Artoisenet:2013puc} is used to generate the \ttHodd\ events. The generated events are interfaced with the Pythia 8.235~\cite{Sjostrand:2014zea} for parton shower. We do not emulate detector effects as the salient features of the problem already present from the parton shower and hadronization. The generated hadrons are used to construct anti-$k_t$~\cite{Cacciari:2008gp} $R=0.4$ jets using \textsc{FastJet} 3.3.2~\cite{Cacciari:2011ma,Cacciari:2005hq}. 

Jets are required to have $|\eta| \leq 2.5$ and $p_T \geq 25$~GeV, while leptons are required to have $|\eta| \leq 2.5$ and $p_T \geq 10$~GeV. A jet is removed if its distance\footnote{$\Delta R$ is defined as $\sqrt{\Delta\eta ^2 + \Delta\phi^2}$, where $\Delta\eta$ is the difference of two particles in pseudorapidity and $\Delta\phi$ is the difference in azimuthal angle.} in $\Delta R$ with a photon or a lepton is less than 0.4. Jets that are $\Delta R$ matched to $b$-quarks at the parton level are labeled as $b$-jets; this label is removed randomly for 30\% of the $b$-jets, to mimic the inefficiency of a realistic $b$-tagging~\cite{ATLAS:2019bwq,CMS:2017wtu}. We further apply a preselection on the testing set of $N_\mathrm{bjet} > 0$, and $(N_\mathrm{jet} \geq 3$ and $N_\mathrm{lepton} = 0)$ or $N_\mathrm{lepton} > 0.$, to mimic realistic data analysis requirements. The \ttH\ and \tttt\ samples are each divided to training, validation, and testing sets, corresponding to a split of 75\%:12.5\%:12.5\%. The other samples ($t\bar{t}$, $t\bar{t}W$, and \ttHodd) are used only for testing. While a single model can be trained to learn from a mixture of processes such as \ttH\ and \tttt\ for greater generality, we leave this exciting direction to future work.

As we compare the performance of CPT to that of a conventional approach, we refer to top quarks that can be matched to a triplet of jets as ``truth-matched'' and those that cannot as ``unmatched''. Specifically, a top quark is considered as ``truth-matched'' if decays hadronically and each of the three quarks originating from its decay is matched ($\Delta R < 0.4$) to exactly one jet. According to this definition, semi-leptonically decaying tops are always unmatched, which is motivated by the fact that we can't physically detect its neutrino (at best we can estimate its kinematics such as $p_\mathrm{T}$). The vast majority (e.g., 76\% for \ttH) of tops are unmatched, and therefore can't be fully reconstructed due to incomplete information about their decay products. For events passing the preselection, the fraction of hadronically decaying top quarks that can be truth-matched is \textcolor{black}{36\% for \ttH, 37\% for \ttbar, 38\% for \ttW, and 38\% for \tttt.}

\section{Performance} 
\label{sec:perf}

We study three different performance aspects of CPT. First, we evaluate the resolution of the predictions of individual top quark kinematic variables. 
Second, we compare the correlation between the predicted variables to the correlations between the true top quark properties. Finally, we assess the model dependence of CPT by applying the model trained on \ttH\ events to alternative processes. We study these metrics inclusively for events passing the preselection, and we also break down the performance for top quarks where a matching triplet of jets can be identified using truth information and for top quarks where no matching triplet of jets can be identified. For the former case, we also compare CPT prediction with the calculation from a triplet-based reconstruction method. The latter scenario corresponds to the case where the conventional triplet-based reconstruction method does not apply.

\begin{figure}[H] 
\centering
\includegraphics[width=0.5\textwidth]{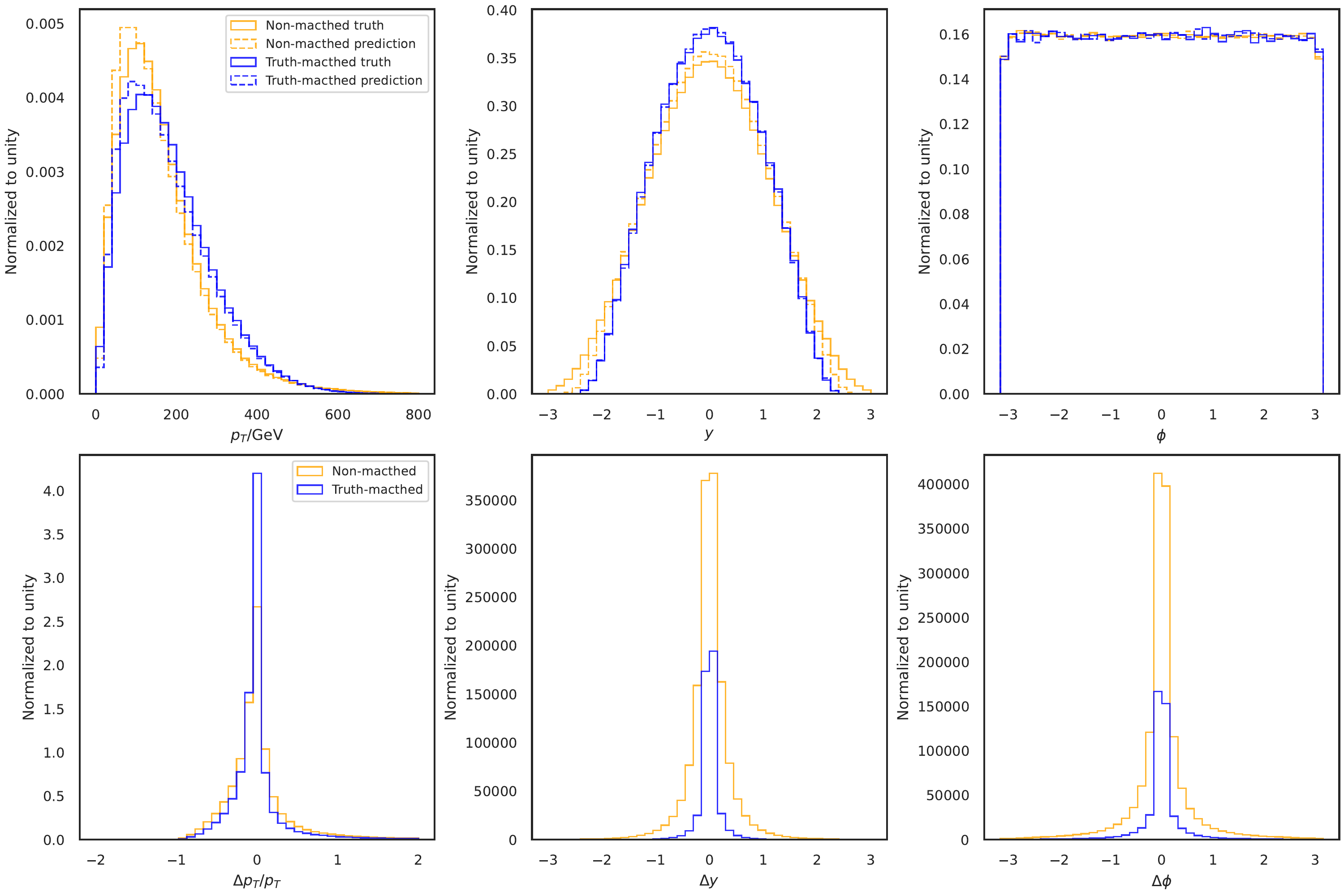}
\caption{Top row: Distributions of truth and predicted top quark four-momentum components, \pt, $y$, and $\phi$ from the \ttH\ sample. Bottom row: The distributions of dimensionless errors $\Delta p_\mathrm{T} / p_\mathrm{T}, \Delta y$, and $\Delta \phi,$ where $\Delta$ means prediction minus truth. The area under each histogram is normalized to unity.
As expected, CPT's performance is worse for unmatched tops due to incomplete information. Over all tops (truth-matched and unmatched) in the test set, the median values of $\Delta p_\mathrm{T} / p_\mathrm{T}, \Delta y$, and $\Delta \phi,$ are $-0.02, 0.002$ and $-0.002,$ showing that there is no significant bias in CPT's prediction.
}
\label{fig:ttHprediction}
\FloatBarrier
\end{figure}  

\noindent\textbf{Resolution:} 
Figure~\ref{fig:ttHprediction} shows the predicted and truth variable distributions for \pt, $\eta$, $\phi$ of the top quarks in the \ttH\ sample. To quantify the resolution, we calculate the width of $\Delta p_\mathrm{T} / p_\mathrm{T,truth }$, $\Delta\eta$ and $\Delta\phi,$ the model's prediction error for the three variables (relative error for $p_\mathrm{T}$). The width is quantified using half of the 68\% inter-quantile range, which corresponds to one standard deviation in the Gaussian case. The top quark mass is part of the four-momentum prediction, but we do not show it here as it is nearly a delta function. Since the model predicts the four-momenta of two top quarks, the predicted top quarks are matched to truth top quarks during the resolution calculation to minimize the sum of $\Delta R$ between all matched pairs. Table~\ref{tab:process_dependence} summarizes the prediction resolutions for all top quarks in the predicted \ttH\ events, separated into ``truth-matched'' top quarks and ``unmatched'' top quarks. As expected, CPT's performance is worse for unmatched tops due to incomplete information. Over all tops (truth-matched and unmatched) in the test set, the median values of $\Delta p_\mathrm{T} / p_\mathrm{T}, \Delta y$, and $\Delta \phi,$ are $-0.02, 0.002$ and $-0.002,$ showing that there is no significant statistical bias in CPT's prediction.

\noindent\textbf{Relative performance:} 
 The model prediction resolutions are compared to the intrinsic resolutions of reconstructing top quarks using jet-triplets. The intrinsic resolutions are calculated from truth-matched triplets of jets, where the four-momenta of the truth-matched jet-triplet are considered as the predictions. In this case, the resolution arises from the effects of quark hadronization and jet reconstruction. For truth-matched top quarks, the ratio of the prediction resolution from CPT to the intrinsic resolution is 1.5 for \pt, 2.3 for the rapidity $y$, and 2.0 for the azimuthal angle $\phi$. 
 
To compare CPT with a strong baseline, we also evaluate a triplet-based reconstruction method, where a neural network is trained to identify the triplet associated with each top quark. The baseline resolutions have prediction-to-intrinsic ratios of 2.2 for \pt, 2.8 for $y$, and 3.1 for $\phi$. Therefore, even when evaluated on truth-matched top quarks, CPT achieves significantly better resolution than the triplet-based method. The comparison is visualized in Figure~\ref{fig:cpt_vs_baseline}. Details on the baseline implementation is available in Appendix~\ref{app:baseline}.

 \begin{figure}[H] 
\centering
\includegraphics[width=0.4\textwidth]{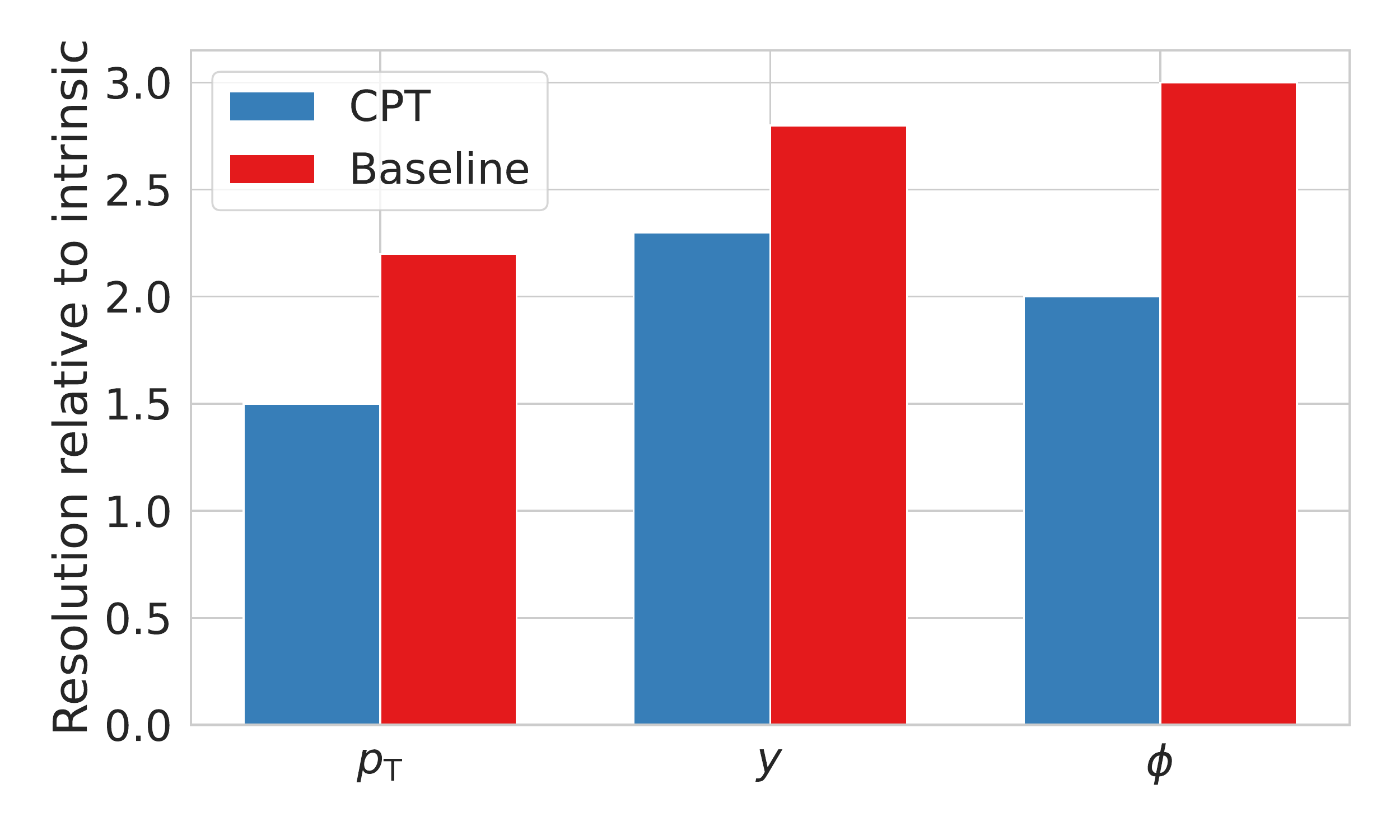}
\caption{Resolution (smaller means better) achieved by CPT and the triplet-based reconstruction (baseline) normalized by the intrinsic resolution arising from effects of quark hadronization and jet reconstruction, evaluated on truth-matched tops in \ttH\ events. CPT achieves significantly better resolution than the reconstruction-based approach.
}
\label{fig:cpt_vs_baseline}
\FloatBarrier
\end{figure}

In the preselected \ttH\ events, 76\%
of the top quarks are unmatched. Specifically, 43\% out of the total 67\% of tops that decay hadronically don't have a matching triplet and 33\% of all tops decay semi-leptonically. For these unmatched top quarks, CPT achieves a prediction-to-intrinsic resolution ratio of 2.5 for \pt, 6.5 for $y$, and 3.6 for $\phi$. Due to incomplete information about the tops' decay products, CPT's performance degrades as expected for unmatched top quarks, though the absolute resolutions remain below 30\%. Note these top quarks cannot otherwise be fully reconstructed using reconstruction-based alternatives due to incomplete information about their decay products. While there exist procedures to approximately recover some of the missing information, such as the neutrino kinematics, combining these additional estimators with a reconstruction-based method to handle unmatched tops introduces additional complexity and sources of error and it's highly unlikely that the resulting approach will outperform a regression model.

\begin{table}[H]
\centering
\caption{Summary of resolutions of top quark four-momentum components in various scenarios for \ttH, \ttbar\, and \ttW processes. 
}
\label{tab:process_dependence}
\begin{tabular}{lrrrr}
\toprule
                & &  ${\sigma_{p_{\mathrm{T}}}}$  & $\sigma_{y}$ & $\sigma_{\phi}$  \\
\midrule

\multirow{3}{4em}{\ttH}&Intrinsic & 0.10 &                          0.04 &                          0.07  \\
                       &Truth-matched & 0.15 & 0.09 & 0.14 \\
                       &Unmatched & 0.27 & 0.25 & 0.26 \\
\midrule
\multirow{3}{4em}{\ttbar}&Intrinsic & 0.11 &                          0.04 &                          0.09\\
                       &Truth-matched & 0.19 &                          0.11 &                          0.20\\
                       &Unmatched & 0.31 &                          0.32 &                          0.37  \\     
\midrule
\multirow{3}{4em}{\ttW}&Intrinsic & 0.12 &                          0.04 &                          0.08  \\
                       &Truth-matched & 0.27 & 0.15 & 0.28 \\
                       &Unmatched & 0.45 & 0.36 & 0.50 \\
       
\bottomrule
\bottomrule
\end{tabular}
\end{table}
\FloatBarrier

\noindent\textbf{Correlation:}
Between the six variables of interest, only three pairs of variables have a linear correlation beyond 5\% in the truth sample. These correlations are $74\%$ for (\ptone,\pttwo), $50\%$ for (\etaone,\etatwo), and $-31\%$ for (\phione,\phitwo). The corresponding correlations observed in the \model\ prediction are $75\%$ for (\ptone,\pttwo), $43\%$ (\etaone,\etatwo), and $-34\%$ for (\phione,\phitwo). The correlation between top quarks is well-reproduced in CPT's predictions.

\begin{figure}
    \centering
    \includegraphics[width=0.9\columnwidth]{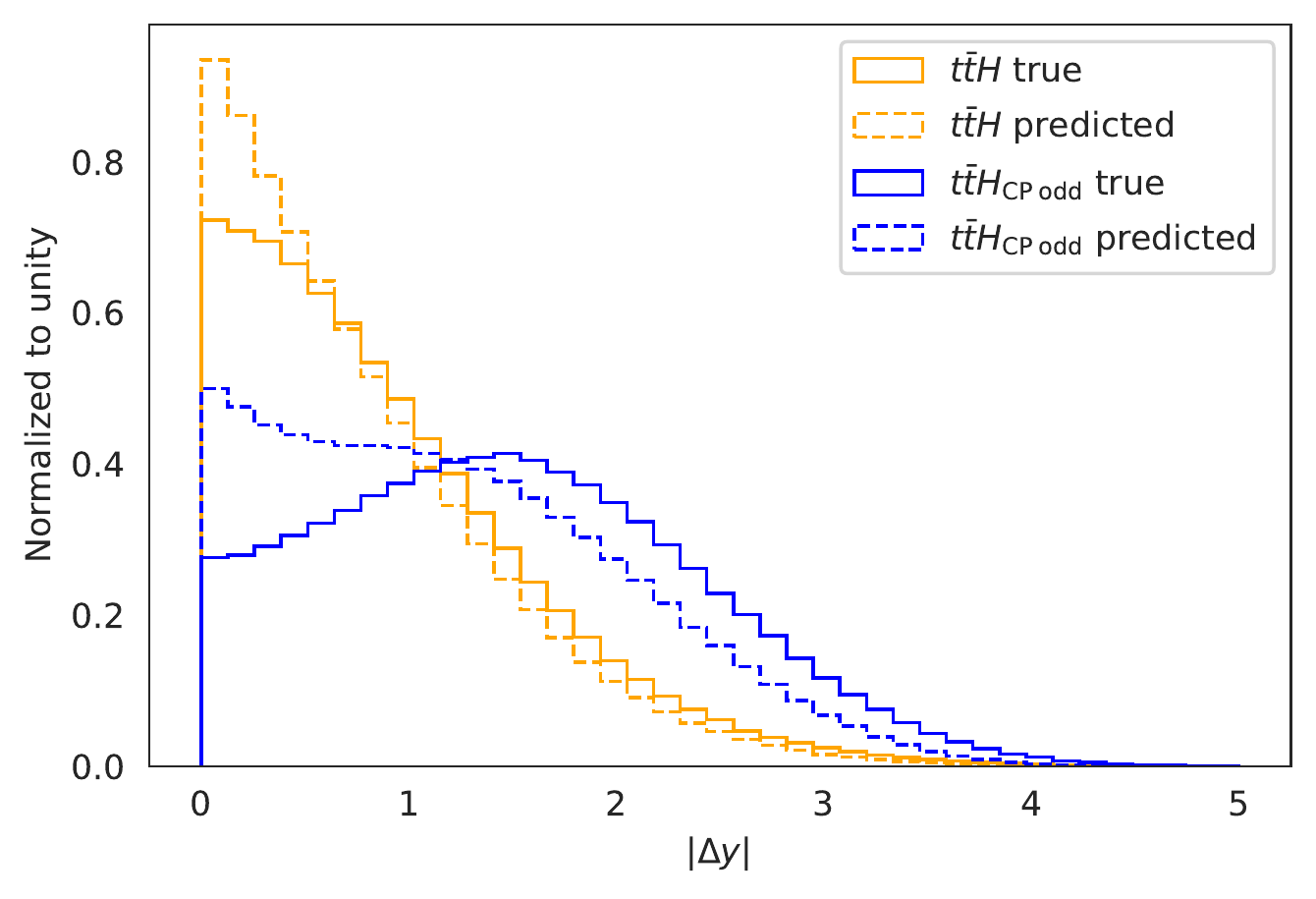}
    \includegraphics[width=0.9\columnwidth]{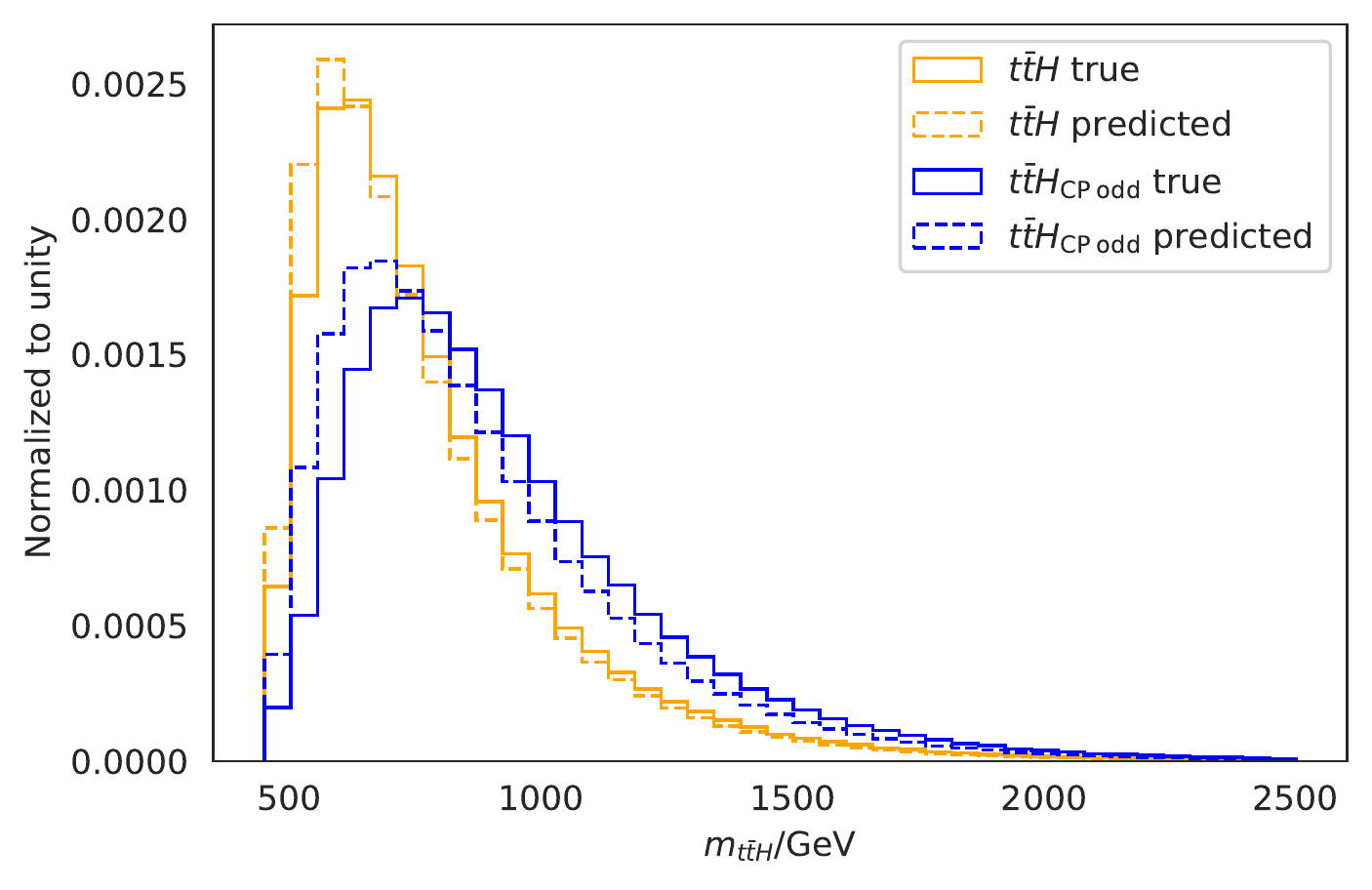}
    \caption{Predicted and truth distributions for system-level observables $|\Delta y|$ (top) and $m_{t\bar{t}H}$ (bottom) in the \ttH\ sample (orange) and \ttHodd\ sample (blue). $|\Delta y|$ is the absolute difference between the rapidities of two tops, and $m_{t\bar{t}H}$ is the invariant mass of the $t\bar{t}H$ system, where the Higgs 4-momentum is taken to be its ground-truth value. The area under each histogram is normalized to unity. As CPT is not trained on the \ttHodd\ sample, its prediction for \ttHodd\ events are worse as expected.
    }
    \label{fig:di-top-observables}
\end{figure}

\vspace{5mm}
\noindent\textbf{Process dependence:}
We assess the process dependence of CPT by applying the model trained with \ttH\ to \ttW, \ttbar\, and \ttHodd\ events, respectively. Table~\ref{tab:process_dependence} compares the intrinsic and prediction resolutions between \ttH, \ttW, and \ttbar\ processes. CPT trained exclusively on the \ttH\ sample can be applied without any retraining to yield a similar level of performance for \ttbar\ events. This level of generalization is not trivial since these two processes induce different statistics in the final state objects and top quarks. The \ttW\ events constitute a much more challenging test set since additional jets, leptons, and neutrinos are produced from the $W$ decay which introduces more complex correlations among the objects that are not present in CPT's training set. Consequently, CPT yields a larger resolution on the \ttW\ test set. The process-dependence can be mitigated by a number of strategies, such as training CPT with a more representative sample or possibly active decorrelation strategies~\cite{Louppe:2016ylz,Stevens:2013dya,Shimmin:2017mfk,Bradshaw:2019ipy,ATL-PHYS-PUB-2018-014,DiscoFever,Xia:2018kgd,Englert:2018cfo,Wunsch:2019qbo,Rogozhnikov:2014zea,10.1088/2632-2153/ab9023,clavijo2020adversarial,Kitouni:2020xgb,Dolan:2021pml}, which we defer to future studies. Figure~\ref{fig:di-top-observables} shows distributions of  the system-level observables constructed from individual top quark four-momenta for \ttH\ and \ttHodd\ samples. A reasonable agreement between the predictions and ground truth properties is observed for these observables, indicating CPT captures the subtle difference in the kinematics between the two processes and reproduces correlation in the four-momentum between the two top quarks. The agreement can be improved by applying preselection such as the requirement of at least one truth-matched top. Importantly, although the model prediction is not perfect, the separation between \ttH\ and \ttHodd\ events is preserved by CPT predictions, showing the promise of applying CPT to produce discriminating kinematic variables. \\

\noindent\textbf{High multiplicity final state:}
CPT can predict the four-momenta of an arbitrary (fixed) number of top quarks in a collision event. We test the prediction ability of CPT in the extreme case at the LHC where four top quarks are produced in the same event. We configure CPT to predict the four-momenta of four top quarks and train it with the \tttt\ sample described in Section~\ref{sec:data}. Table~\ref{tab:four-top} shows the intrinsic and prediction resolutions from this test. Compared to the prediction for the \ttH\ sample, the prediction for \tttt\ is worse. However, the intrinsic resolution in the \tttt\ sample is also worse than that in the \ttH\ sample, suggesting that the top quarks in \tttt\ events are inherently more complex and challenging to reconstruct. We expect the gap between the intrinsic and CPT's resolution can be reduced by further architectural improvements and more training data. We stress that the exploding combinatorics in \tttt\ events render reconstruction-based methods prohibitively expensive to be successfully applied in this setting, whereas we can easily apply CPT without any modification. To predict top quarks' kinematics from $N$ jets, a standard reconstruction-based method has a super-exponential computational complexity of $O(N!),$ the number of all possible permutations within $N$ objects, while CPT only has a polynomial complexity of $O(N^2)$ since the attention mechanism only involves pairwise interactions among the objects.

\begin{table}[H]
\centering
\caption{Summary of resolutions of top quark four-momentum components in various scenarios in the \tttt\ sample. 
}
\label{tab:four-top}
\begin{tabular}{lrrrr}
\toprule
                & ${\sigma_{p_{\mathrm{T}}}}$ & $\sigma_{y}$ & $\sigma_{\phi}$  \\
\midrule
Intrinsic  & 0.19 &                          0.05 &                          0.09\\

Truth-matched & 0.29 &                          0.16 &                          0.24\\
Unmatched & 0.42 &                          0.32 &                          0.36\\
\bottomrule
\bottomrule
\end{tabular}
\end{table}
\FloatBarrier

\section{Ablation studies}
\label{sec:ablat}
We demonstrate the effects of removing important components of CPT to show how they contribute to the final performance. All comparisons are done on the \ttH\ dataset. Resolutions are reported on all top quarks passing the preselection, regardless of truth-matching status.\\

\noindent\textbf{Attention mechanism:}
The attention mechanism is an important part of the model as it allows the model to selectively focus on a subset of the final state objects in determining the four-momentum of each top quark. We demonstrate its benefit by training an otherwise identical model except with all attention weights set to a constant $\frac{1}{N_\mathrm{in}},$ where $N_\mathrm{in}$ is the number of final state objects in the event. Comparisons between the resolution achieved by this model and the nominal model is shown in Table~\ref{tab:no_attention}. We observe the model with uniform attention achieves worse resolutions, which demonstrates the benefit of the attention mechanism.

\begin{table}[H]
\centering
\caption{Comparison of resolutions of top quark four-momentum components in the \ttH\ sample achieved by CPT and its variant applying uniform-attention for each final state object. }
\label{tab:no_attention}
\begin{tabular}{lrrrr}
\toprule
                & ${\sigma_{p_{\mathrm{T}}}}$ & $\sigma_{y}$ & $\sigma_{\phi}$  \\
\midrule
CPT & 0.24 & 0.21 & 0.23 \\
CPT (uniform attention) & 0.27 & 0.23 & 0.28 \\
\bottomrule
\bottomrule
\end{tabular}
\end{table}

\begin{figure}[H]     
\centering     
\includegraphics[width=1\columnwidth]{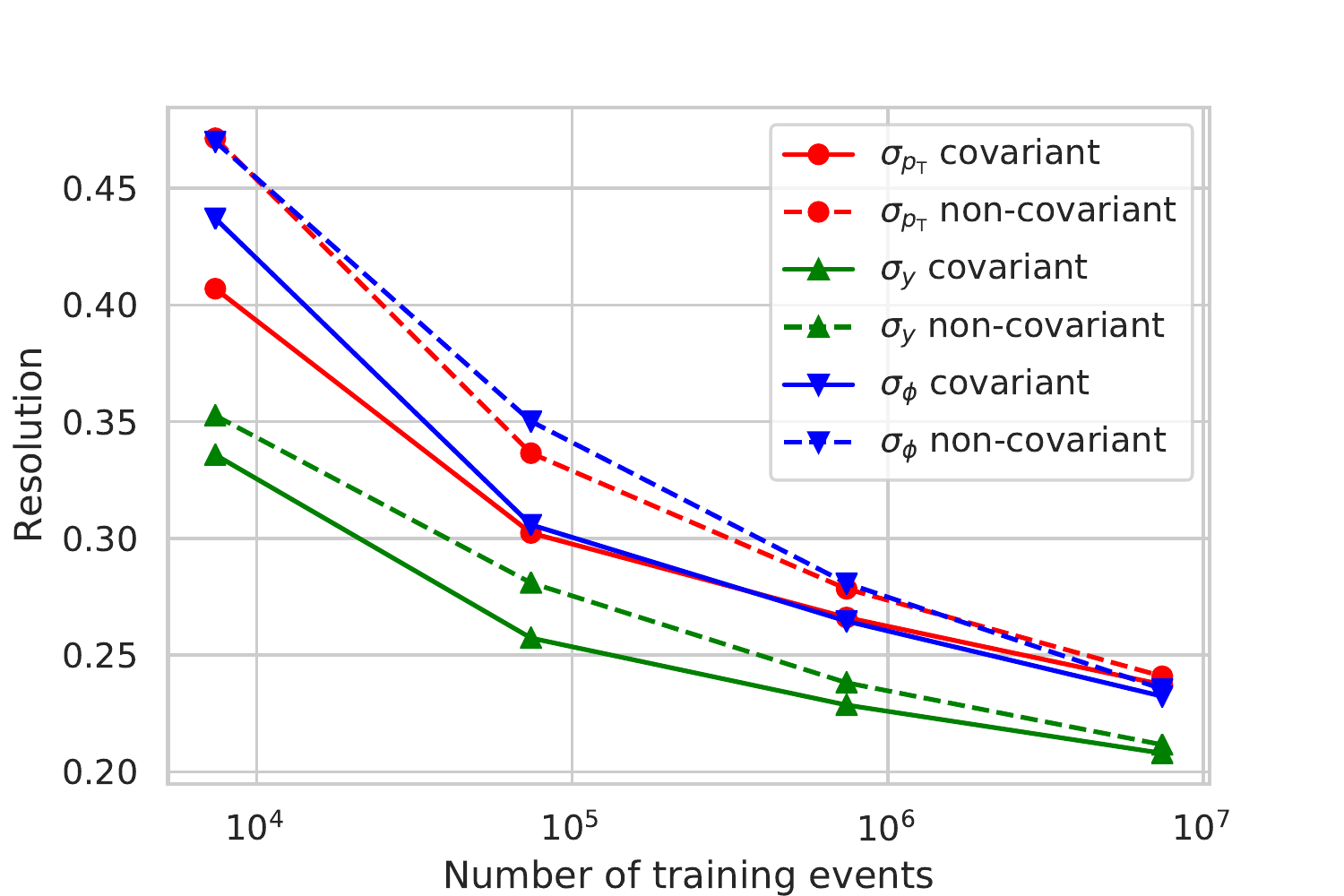}
\caption{Resolution on in \ttH\ sample achieved by using the covariant attention and non-covariant attention. The covariant attention offers clear benefit particularly in the low-data regime. 
}
\label{fig:cov_vs_non_cov} 
\end{figure}

\noindent\textbf{Covariant attention:} CPT employes a covariant attention mechanism to exploit the symmetries in collision data. When the covariant attention is replaced by a regular attention mechanism which does not guarantee covariance, we observe increasing degradation in performance as the size of the training sample becomes smaller. Figure~\ref{fig:cov_vs_non_cov} compares the resolutions achieved by CPT and its variant using a regular attention mechanism, as a function of the number of training events. For example, the increase in \pt\ resolution can be as large as \textcolor{black}{16\% when only 0.1\% of the events in the nominal training sample is used.} This shows that the covariant attention enables CPT to be more data-efficient and provide more accurate predictions in the low-data regime compared to non-covariant models.

\noindent\textbf{Alternative architectures:}
Finally, we compare with two alternative permutation-invariant architectures, Graph Convolutional Networks~\cite{kipf2016semisupervised} and DeepSets~\cite{zaheer2017}.
Applied to this task, Graph Convolutional Networks (GCNs) use graph convolutions to process information in the final state objects represented as a complete graph, while DeepSets uses a fully connected neural network encoder to learn the feature vector of each final state object individually. In both cases, the feature vectors of all final state objects are then summed and fed into a fully connected neural network to predict the top quark four-momenta.  The \model\ mainly differs from these two architectures by utilizing an attention mechanism, implementing partial Lorentz covariance, and using a decoder module. We use six graph convolutional layers and six encoder layers for the GCN and the DeepSet models, and a feature dimension of 128 for both. A comparison of resolutions between the models is shown in Table~\ref{tab:baseline}. CPT significantly outperforms the other two methods, showing its outstanding effectiveness on this task.  We did not perform extensive hyperparameter optimizations for any of the three architectures.  However, we hypothesize that the performance ordering would persist after such an optimization given the magnitude of the observed differences.  We defer this study to future work.

\begin{table}[H]
\centering
\caption{Comparison of resolutions of top quark four-momentum components in the \ttH\ sample achieved by CPT, GCN, and DeepSets. 
\label{tab:baseline}
}
\begin{tabular}{lrrrr}
\toprule
                & ${\sigma_{p_{\mathrm{T}}}}$ & $\sigma_{y}$ & $\sigma_{\phi}$  \\
\midrule
CPT & 0.24 & 0.21 & 0.23 \\
GCN & 0.38 & 0.35 & 0.42 \\
DeepSets & 0.36 & 0.32 & 0.36 \\
\bottomrule
\bottomrule
\end{tabular}
\end{table}

\section{Conclusion}
\label{sec:concl}
In this paper, we propose a new machine learning-enabled approach to determining top quark kinematic properties by processing the full event information holistically. Our approach offers three major advantages compared to conventional approaches. First, we no longer deal with the conceptually ill-defined jet-triplet identification process. Second, we can account for noisy or missing observations due to limited detector acceptance, inefficiency, and resolution, as the regression model can learn such effects from simulations. Third, the holistic processing of the event final state offers a unified approach to determine the top quark properties for both the hadronic and semi-leptonic top quark decays, which simplifies the analysis workflow. Finally, our approach has a runtime polynomial in the number of final state objects as opposed to super-exponential for reconstruction-based approaches which need to consider all possible permutations, making ours the first tractable method for processes with high multiplicity final state such as \tttt.

To realize this holistic approach to predicting top quark kinematic properties, we propose the Covariant Particle Transformer (CPT). CPT takes as input properties of the final state objects in a collision event and outputs predictions for the top quark kinematic properties. Using a novel covariant attention mechanism, CPT prediction is invariant under permutation of the inputs and covariant under rotation and/or boosts of the event along the beamline. CPT can recover \textcolor{black}{76\% (75\%)} of the top quarks produced in the \ttH\ (\tttt) events that cannot be truth matched to a jet-triplet and thus not fully reconstructable by conventional methods. For \ttH\ events, CPT achieves a resolution close to the intrinsic resolution of jet-triplet and outperforms a carefully tuned triplet-based top reconstruction method on top quarks that can be matched to a jet-triplet. In addition, we demonstrate that CPT can generalize to top production processes not seen during training, though its performance degrades as the test process becomes more complex and distinct from the training process. Finally, we demonstrate that by building Lorentz covariance into CPT, it achieves higher data efficiency and outperforms the non-covariant alternative when the training set is small.

In the future, it may be possible to improve and extend CPT.  CPT training uses simulation to learn to invert parton shower and hadronization (and in the future, detector effects).  Training strategies that rely less on parton shower and hadronization simulations like those in Ref.~\cite{Howard:2021pos} may be able to improve the robustness of CPT.  Furthermore, as a direct regression approach, CPT is prior dependent.  A variety of domain adaptation and other strategies may be able to further improve the resilience of CPT.  It may also be possible to include lower-level, higher-dimensional inputs directly into CPT instead of first clustering jets.

As it uses a generic representation for collision events as sets of particles, CPT can be directly applied to predict kinematic properties of other heavy decaying particles, such as the $W$, $Z$, and Higgs boson, and potential heavy particles beyond the SM. The predicted kinematics of these heavy decaying particles can be used to construct discriminating variables for searches or observables for differential cross-section measurements. The ability to predict properties of heavy decaying particles through a holistic analysis of the collision event can enable measurements that otherwise suffer extreme inefficiencies using conventional reconstruction methods. \\

\noindent\textbf{Note added:} While this paper was being finalized, we became aware of Ref.~\cite{Gong:2022lye}, which proposes another Lorentz equivariant architecture.  In contrast to that paper, we integrate Lorentz covariance with the Transformer, a state-of-the-art neural network architecture that revolutionized many areas of machine learning applications such as natural language processing, computer vision, and protein folding. We have also considered a completely different application: namely regression instead of classification, where the Lorentz group acts non-trivially (not an identity) on the target variables.

\section*{Acknowledgements}
This work is supported by the U.S.~Department of Energy, Office of Science under contract DE-AC02-05CH11231. H.W.'s work is partly supported by the U.S. National Science Foundation under the Award No. 2046280.

\appendix
\section{CPT Implementation}
\label{app:cpt_implementation}
\subsection{Attention mechanism}
Attention mechanisms are a way to update a vector of $n$ features $\{x_i\}_{i=1}^{n}$,  given a context $\{c_j\}_{j=1}^{m}$. Learnable query, key, and value matrices $\{W_Q, W_K, W_V\}$ are used to generate $d$-dimensional query, key, and value vectors $\{q_i\}_{i=1}^{n},$ $\{k_j\}_{j=1}^{m},$ and $\{v_j\}_{j=1}^{m},$ via
\begin{align}
    q_i &= W_Q x_i \\
    k_{j} &= W_K c_j,\\
    v_{j} &= W_V c_j.
\end{align}
The inner product between $q_i^\top$ and $k_j$ is used to compute the attention weights $\alpha_{ij}$ through
\begin{align}
    \alpha_{ij} = \frac{\exp(q_i^\top k_j / \sqrt{d})}{\sum_j \exp(q_i^\top k_j / \sqrt{d})},
\end{align}
where the $\sqrt{d}$ is a normalization factor. A weighted sum of the value vectors are then used to compute update vectors $\{m_i\}_{i=1}^{n},$
\begin{align}
    m_i = \sum_j \alpha_{ij} v_j,
\end{align}
which is then used to update $x_i$ by, for example, addition $x'_i = x_i + m_i$. Intuitively, the attention weights $\alpha_{ij}$ represent how important the information contained in $c_j$ is to $x_i.$ When the context $\{c_j\}$ is simply $\{x_i\},$ this is termed as self-attention, otherwise cross-attention. It is common to use a slight extension of the method above, called Multi-headed attention, where $H$ different query, key, and value matrices $\{(W^h_Q, W^h_K, W^h_V)\}_{h=1}^{H}$ are learnt. Each head follow the above procedure to independently produce attention weights $\{a_{ij}^h\}_{ijh}$ and then update vectors $\{m^h_i\}_{i=1,h=1}^{n, H}.$ The $H$ update vectors $\{m^h_i\}_{h=1}^{H}$ received by each $x_i$ are concatenated to produce a final update vector
\begin{align}
    m_i = \bigoplus_{h=1}^{H} m^h_i,
\end{align}
which is then used to update $x_i$ as before.

\subsection{Particle representation}
We represent each particle with a feature vector $h_i,$ and $h_i=(x_i, \omega_i)$ consists of an invariant feature vector $x_i,$ and a covariant featrue vector $\omega_i.$ $x_i$ is an invariant quantity under a rotation and boost along the beamline, while $\omega_i=(\eta_i, \cos(\phi_i), \sin(\phi_i))$ represnets the flight direction of the object and is a covariant quantity. As input to the \model, $x_i=(p_{\mathrm{T},i}, m_i, \mathrm{id})$ where $\mathrm{id}$ is a one-hot vector indicating particle identity. The model learns to update these feature vectors while maintaining their invariance/covariance property through the \attn.

\subsection{Covariant attention}
To update the learned feature vectors of each object in the event, we use covariant attention, an extenstion of the regular attention mechanism to process kinematics information and gaurantee covariance properties of the predictions. In general, covariant attention updates feature vectors $\{h_i\}$ of a subset of the objects in the event using feature vectors $\{h_j\}$ of a (potentially different) subset as context. First, it computes the flight direction of each context object as viewed in $i$'s frame: $\omega_{ij} = (\eta_j - \eta_i, \cos(\phi_j - \phi_i), \sin(\phi_j - \phi_i),$ which is invariant under longitudinal boosts and azimuthal rotations. Then it computes the $d$-dimensional query, keys, and value vectors as follows
\begin{align}
    \hat{x}_i &= \mathrm{LayerNorm}(x_i), \\
    v_{ij} &= W_V(\hat{x}_j + \mathrm{MLP}(\omega_{ij})), \\
    k_{ij} &= W_K(\hat{x}_j + \mathrm{MLP}(\omega_{ij})),\\
    q_i &= W_Q \hat{x}_i
\end{align}
where $W_V, W_K, W_Q$ are learned matrices and $\mathrm{MLP}$ is a Multilayer Perceptron. The inner products between $q_i$ and $k_{ij}$ are then went through softmax operator so as to weight the value vectors. The weighted sum produces an aggregated message vetor $m^x_i$ which is added to $x_i$:
\begin{align}
    \alpha_{ij} &= \frac{\exp(q_i^\top k_{ij} / \sqrt{d})}{\sum_j \exp(q_i^\top k_{ij} / \sqrt{d})}, \\
    \tilde{m}^x_i &= \sum_j \alpha_{ij} v_{ij}, \\
    m^x_i &= \sigma(\mathrm{Linear}(x_i, \tilde{m}^x_i)) \odot \tilde{m}^x_i, \\
    x'_i &= x_i + m^x_i,
\end{align}
where $\sigma$ is the sigmoid function and $\odot$ denotes elementwise (Hadamard) product. Gating is applied to the attention weights following the Gated Attention Network~\cite{zhang2018gaan}. A multi-headed version of covariant attention can be constructed in the same way as in regular attention, and is omitted here. $x'_i$ is then passed through a feed-forward network as done in the original transformer. When it is desirable to also update the covariant feature $\omega_i,$ we produce another update vector $m^\omega_i$ from $m^x_i$ via
\begin{align}
    \tilde{m}^\omega_i &= \mathrm{MLP}(m^x_i) \\
    m^\omega_i &= \sigma(\mathrm{Linear}(x_i, \tilde{m}^\omega_i)) \odot \tilde{m}^\omega_i, \\
\end{align}
where $m^\omega_i$ is a three dimensional vector. Its first component is used as a boost with rapidity $\delta\eta_i,$ while its last two components $v_i$ converted to a rotation matrix $R\qty(v_i),$ which is used to rotate the azimuthal angle $\phi_i:$
\begin{align}
    \eta'_i &= \eta_i + \delta \eta_i \\
    \begin{pmatrix}
    \cos(\phi'_i)\\
    \sin(\phi'_i)
    \end{pmatrix} &= R\qty(v_i) \begin{pmatrix}
    \cos(\phi_i)\\
    \sin(\phi_i)
    \end{pmatrix}\\
    \omega'_i &=(\eta'_i, \cos(\phi'_i), \sin(\phi'_i)).
\end{align}
where $R(v_i)$ is obtained as follows
\begin{align}
    u_i &= v_i + (1, 0), \label{eq:rotate_phi}\\
    w_i &= \frac{u_i}{\norm{u_i}} = (\cos(\theta_i), \sin(\theta_i)),\\
    R\qty(v_i) =& \begin{pmatrix}
    \cos(\theta_i) & -\sin(\theta_i)\\
    \sin(\theta_i) & \cos(\theta_i)
    \end{pmatrix},
\end{align}
where we added $(1, 0)$ to $v_i$ to bias the rotation matrix to an identity for stability. The covariance of $\{\omega'_i\}$ follows from the fact that only invariant information is used to construct its update, and prior to the update, $\{\omega_i\}$ are themselves covariant. An inductive argument establishes the end-to-end covariance of compositions of covariant attention updates. We denote the above covariant attention update as $h_i \leftarrow \A^{x\omega}_{x\omega}(h_i, \{h_j\})$ where the subscript indicates that it makes use of both the invariant and covariant feature vector, and the superscript indicates that it updates both the invariant and covariant feature vector. The following variants are used to build the full model:
\begin{itemize}
    \item $x_i \leftarrow \A^{x}_{x\omega}(h_i, \{h_j\}):$ the covariant feature vector is not updated
    \item $x_i \leftarrow \A^{x}_{x}(x_i, \{x_j\}):$ the covariant feature vector is not updated nor used to construct the key and value vectors. This reduces to the regular attention mechanism.
\end{itemize}

\subsection{Encoder}
The encoder uses six layers of \attn\ to update the input invariant features $x^{\mathrm{in}}_i \leftarrow \A^{x}_{x\omega}(h^{\mathrm{in}}_i, \{h^{\mathrm{in}}_j\})$. The covariant features associated with the input objects $\{\omega^{\mathrm{in}}_i\}$ are not updated.
\subsection{Decoder}
\label{sec:output_init}
\subsubsection{Initialization}
The decoder first initializes the invariant feature vectors associated with the top quarks using the Set2Set module~\cite{vinyals2016order}, which takes in the set $\{x^{\mathrm{in}}_i\}$ and outputs $\{x^{\mathrm{out}}_i\}$, the initial invariant feature vectors of the output objects. The decoder then updates $\{x^{\mathrm{out}}_i\}$ by having each output attends to the input objects, using invariant features only, $x^{\mathrm{out}}_i \leftarrow \A^{x}_{x}(x^\mathrm{out}_i, \{x^{\mathrm{in}}_j\}).$ The attention weights $\alpha_{ij}$ computed in the previous attention update is used to intialize the output covariant feature vectors:
\begin{align}
    \eta^\mathrm{out}_i &= \sum_j \alpha_{ij} \eta^\mathrm{in}_j, \\
    \begin{pmatrix}
    \cos(\phi^\mathrm{out}_i)\\
    \sin(\phi^\mathrm{out}_i)
    \end{pmatrix} &= \frac{\sum_j \alpha_{ij} \begin{pmatrix}
    \cos(\phi^\mathrm{in}_j)\\
    \sin(\phi^\mathrm{in}_j)
    \end{pmatrix}}{\norm{\sum_j \alpha_{ij} \begin{pmatrix}
    \cos(\phi^\mathrm{in}_j)\\
    \sin(\phi^\mathrm{in}_j)
    \end{pmatrix}}}
\end{align}
The covariance of $\eta^\mathrm{out}_i$ follows from the fact that $\sum_j \alpha_{ij} = 1,$ and $\{\eta^\mathrm{in}_j\}$ transforms by an overall additive constant under a boost. The covariance of $\phi^\mathrm{out}_i$ follows from the fact that its unit vector representation is a linear combination of $\qty{\begin{pmatrix}
    \cos(\phi^\mathrm{in}_j)\\
    \sin(\phi^\mathrm{in}_j)
    \end{pmatrix}}_j,$ each of which transform linearly by a rotation. 
\subsubsection{Interleaved covariant cross- and self-attention}
After initialization, the decoder consists of $L_\mathrm{out}=6$ decoder blocks. In each block, the output invariant and covariant feature vectors are updated using two \attn\ layers:
\begin{align}
    h^{\mathrm{out}}_i &\leftarrow \A^{x\omega}_{x\omega}(h^\mathrm{out}_i \{h^{\mathrm{out}}_j\}) \quad\forall i, \\
    h^{\mathrm{out}}_i &\leftarrow \A^{x\omega}_{x\omega}(h^\mathrm{out}_i, \{h^{\mathrm{in}}_j\}) \quad \forall i. \label{eq:cross_attn}
\end{align}
After each decoder block, indexed by $\ell \in \{1, ..., L_\mathrm{out}\},$ an intermediate set of predictions $\{p^\ell_i\}_i$ for the top quark four momenta is constructed as follows
\begin{align}\nonumber
    (p^\ell_{T_i}&/\mathrm{GeV}, \eta^\ell_i, \phi^\ell_i, m^\ell_i/\mathrm{GeV}) \\
    &= (100 (x^\ell_i)_0, \eta_i, \phi_i, 5 (x^\ell_i)_1 + 173),
\end{align}
where $(x^\ell_i)_0, (x^\ell_i)_1$ denotes the first and second entry of the invariant feature vector associated with each top at the $\ell$-th block. The shift and scaling is to keep the feature vectors small and centered to facilitate training.

\subsection{Loss function and optimization details}
\label{sec:loss}

For each event, the main component of loss function is the $L_2$ norm of the difference between the model prediction and ground truth for the top quark four-momenta in $(p_x/100~\mathrm{GeV}, p_y/100~\mathrm{GeV}, \eta, m/5~\mathrm{GeV})$ coordinates, averaged over the $N$ top quarks present in the event:

\begin{equation}
    \mathcal{L}_\mathrm{final} = \frac{1}{N}\sum_{i=1}^{N} \norm{p_i - p^*_i},
\end{equation}
where $\{p_i\}$ are the model predictions at the final decoding block and $\{p^*_i\}$ are the ground truths. We chose this set of coordinates so that each component of the four-momenta has standard deviation of $O(1),$ encouraging the model to pay equal attention to each of them. The $N$ predictions from the model are matched to the $N$ ground truths through a permutation $\pi^*$ that minimizes the average $\Delta R$ between each matched pair:
\begin{equation}
    \label{eq:perm}
    \pi^* = \argmin_{\pi:\mathrm{permutations}} \frac{1}{N}\sum_{i=1}^{N} \sqrt{(\eta_i - \eta^*_i)^2 + (\phi_i - \phi^*_i)^2}.
\end{equation}
We add two auxiliary losses $\mathcal{L}_\mathrm{intermediate}$ and $\mathcal{L}_{\mathrm{unit}-\mathrm{norm}}$ to stabilize training models with many layers. The intermediate loss $\mathcal{L}_\mathrm{intermediate}$ measures the intermediate prediction errors at earlier decoder blocks,

\begin{equation}
    \mathcal{L}_\mathrm{intermediate} = \frac{1}{L_\mathrm{out} - 1}\sum_{\ell=1}^{L_\mathrm{out}-1} \qty(\frac{1}{N}\sum_{i=1}^{N} \norm{p^\ell_i - p^*_i}),
\end{equation}
where $\{p^\ell_i\}_{i=1}^{N}$ are intermediate predictions at the $\ell^\text{th}$ decoder. The unit-norm loss $\mathcal{L}_{\mathrm{unit}-\mathrm{norm}}$ encourages the vectors $u_i$ to have unit-norm before being normalized and converted to rotation matrices in Equation~\ref{eq:rotate_phi} in each decoder block:
\begin{equation}
    \mathcal{L}_{\mathrm{unit}-\mathrm{norm}} = \frac{1}{L_\mathrm{out}}\sum_{\ell=1}^{L_\mathrm{out}} \qty(\frac{1}{N}\sum_{i=1}^{N} \abs{\norm{u^\ell_i} - 1}).
\end{equation}
The two auxiliary losses are inspired by similar auxiliary losses in AlphaFold2~\cite{jumper2021highly}. The total loss is a weighted combination of the above three terms,
\begin{equation}
    \mathcal{L}_\mathrm{total} = \lambda_1 \mathcal{L}_\mathrm{final} + \lambda_2 \mathcal{L}_\mathrm{intermediate} + \lambda_3 \mathcal{L}_{\mathrm{unit}-\mathrm{norm}}.
\end{equation}
We use $\lambda_1=\lambda_2=1,$ and $\lambda_3=0.02$. All models used to report our results are trained using the Lamb optimizer~\cite{conf/iclr/YouLRHKBSDKH20} with a batch size of 256 and a learning rate of $10^{-4}$ for 30 epochs and $10^{-5}$ for another 10 epochs. A weight decay~\cite{loshchilov2017decoupled} of 0.01 is applied. Model from the epoch achieving minimum validation loss is used for final evaluation. This training protocol is sufficient to saturate validation performance for all variants of the model and datasets of various processes and sizes used to present our results.

\section{Baseline}
\label{app:baseline}
We train a neural network to identify triplets of jets that originate from top decays. This task can be formulated as a link prediction problem on a graph, where the nodes are detected jets in an event and any two jets that belong to a triplet are connected by a link. Specifically, every event is represented by a fully-connected graph using the four-momenta and particle types as node features and a graph neural network (GNN) predicts a probability $p_{ij} \in (0, 1)$  that a link exists between jet $i$ and jet $j$ for every pair of jets in the event. The particular architecture we use is the Interaction Network~\cite{battaglia2016interaction}, followed by an MLP applied per edge to output the per-edge probabilities. The GNN is trained to minimize the cross-entropy loss so that $p_{ij}$ is encouraged to be 1 if the jets belong to the same triplet and 0 otherwise. It uses the same training, validation, and test set as used by CPT. We tune the hyperparameters to maximize validation accuracy and settled on 4 Interaction Network blocks, 2 layers and 128 hidden units for all MLPs, Adam optimizer, and a learning rate of 0.001. At test time, we sort all possible links $(i,j)$ by decreasing order in $p_{ij}$ and sequentially form one or two predicted triplets depending on the number of available jets in the event. Each predicted top four-vector is the system four-vector of the predicted triplet. The predicted tops are $\Delta R$ matched to the true tops following the same procedure in CPT defined in Equation~\ref{eq:perm}. We note that this method provides a strong baseline as it uses a neural network architecture that has demonstrated state-of-the-art performance on reasoning about object and relations in a wide range of complex problems such as $N$-body dynamics and estimating physical quantities~\cite{battaglia2016interaction}.

\bibliography{references,HEPML}
\newpage
\appendix

\end{document}